\documentclass{appolb}
\usepackage{epsfig}
\usepackage{subfigure}
\begin{document}
\title{Charge Dependence and Scaling Properties of Dynamical $K/\pi$, $p/\pi$, and $K/p$ Fluctuations from the STAR Experiment%
\thanks{Presented at the International Conference on Strangeness in Quark Matter 2011 (SQM 2011), Krakow, Poland, September 18-24, 2011}%
}
\author{Terence J Tarnowsky (for the STAR Collaboration)
\address{National Superconducting Cyclotron Laboratory, Michigan State University\\ East Lansing, MI, USA}	
}
\maketitle
\begin{abstract}
Dynamical fluctuations in global conserved quantities such as baryon number, strangeness, or charge may be enhanced near a QCD critical point. Charge dependent results from new measurements of dynamical $K/\pi$, $p/\pi$, and $K/p$ ratio fluctuations are presented. The STAR experiment has performed a comprehensive study of the energy dependence of these dynamical fluctuations in Au+Au collisions at the energies $\sqrt{s_{NN}}$ = 7.7-200 GeV using the observable, $\nu_{\rm dyn}$. These results are compared to previous measurements and to theoretical predictions. Various proposed scaling scenarios that attempt to remove the intrinsic volume dependence of $\nu_{\rm dyn}$ and to simplify comparisons between experimental measurements are also considered. Constructing an intensive quantity allows for a direct connection to thermodynamic predictions.
\end{abstract}
  
\section{Introduction}

Fluctuations and correlations are well known signatures of phase transitions. In particular, the quark/gluon to hadronic phase transition may lead to significant fluctuations \cite{Koch1}. In 2010, the Relativistic Heavy Ion Collider (RHIC) began a program to search for the QCD critical point. This involves an ``energy scan'' of Au+Au collisions from top collision energy ($\sqrt{s_{NN}}$ = 200 GeV) down to energies as low as $\sqrt{s_{NN}}$ = 7.7 GeV \cite{STARBES}. This critical point search will make use of the study of correlations and fluctuations, particularly those that could be enhanced during a phase transition that passes close to a critical point.

$\nu_{\rm dyn}$ quantifies deviations in the particle ratios from those expected for an ideal statistical Poissonian distribution \cite{nudyn1, nudyn2}. The definition of $\nu_{\rm dyn,K/\pi}$ (describing fluctuations in the $K/\pi$ ratio) is,
\begin{eqnarray}
\nu_{\rm dyn,K/\pi} = \frac{<N_{K}(N_{K}-1)>}{<N_{K}>^{2}}
+ \frac{<N_{\pi}(N_{\pi}-1)>}{<N_{\pi}>^{2}}
- 2\frac{<N_{K}N_{\pi}>}{<N_{K}><N_{\pi}>}\ .
\label{nudyn}
\end{eqnarray}

A formula similar to (\ref{nudyn}) can be constructed for $p/\pi$ and $K/p$ ratio fluctuations. Additional information about $\nu_{\rm dyn}$ can be found in \cite{nudyn2,starkpiprl,DPF2011_proceeding}. An in-depth study of $K/\pi$ fluctuations in Au+Au collisions at $\sqrt{s_{NN}}$ = 200 and 62.4 GeV was previously carried out by the STAR experiment \cite{starkpiprl}.

%
%
\section{Results and Discussion}

\begin{figure}
\centering
\includegraphics[width=0.70\textwidth]{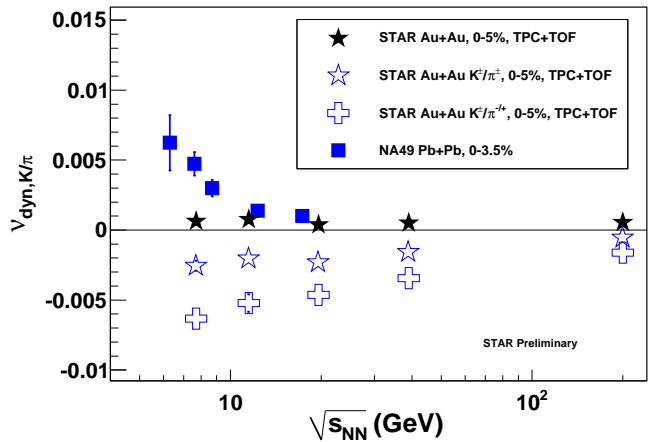}
\caption{Results for the measurement of inclusive charged (closed stars) and average same-sign (open stars) and opposite-sign (open crosses), charge dependent $\nu_{\rm dyn,K/\pi}$.}
\label{nudyn_kpi_excitation}
\end{figure}

\begin{figure}[]
\advance\leftskip-2cm
\subfigure[]{
\includegraphics[width=0.66\textwidth]{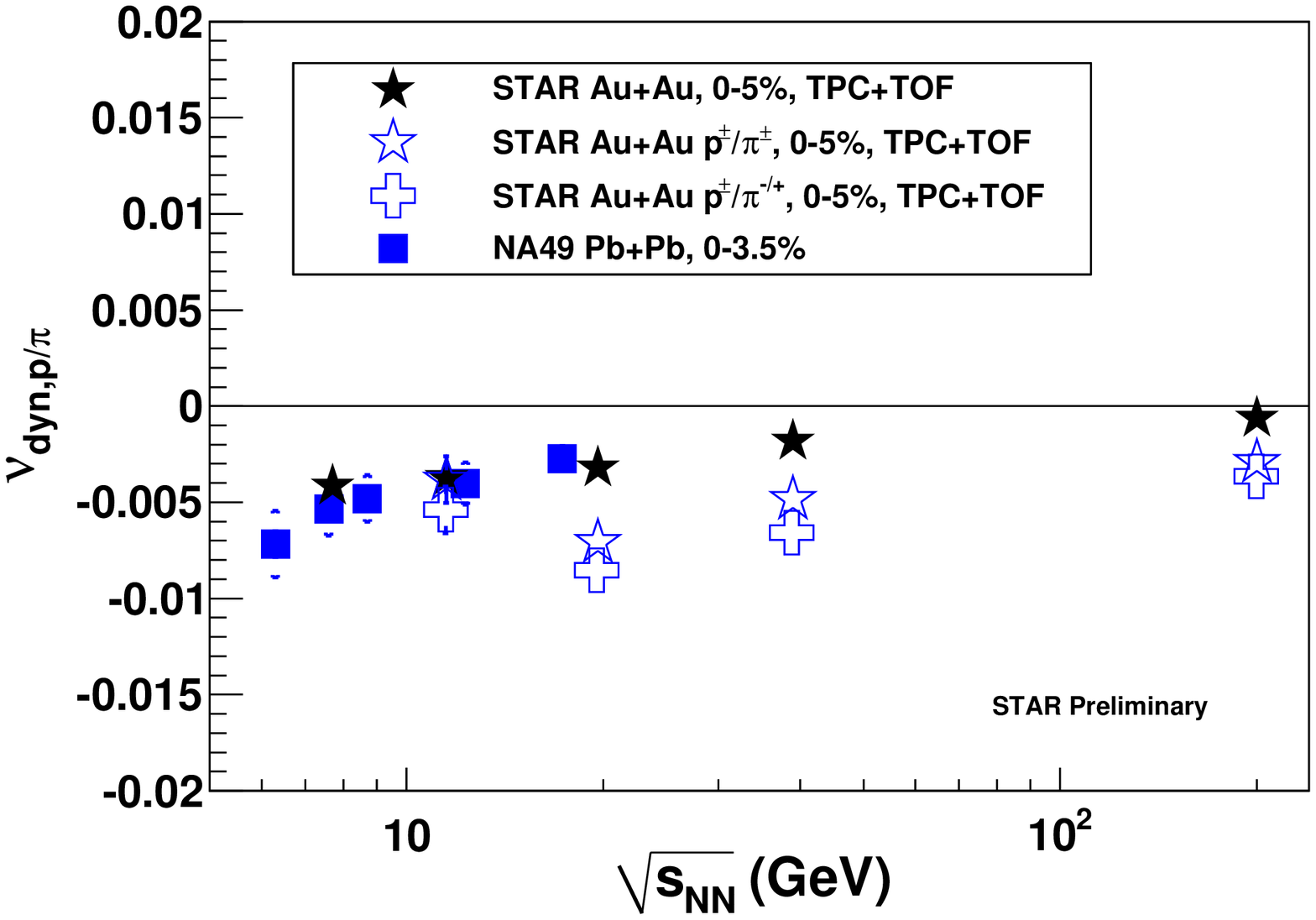} 
\label{nudyn_ppi_excitation}
}
\subfigure[]{
\includegraphics[width=0.66\textwidth]{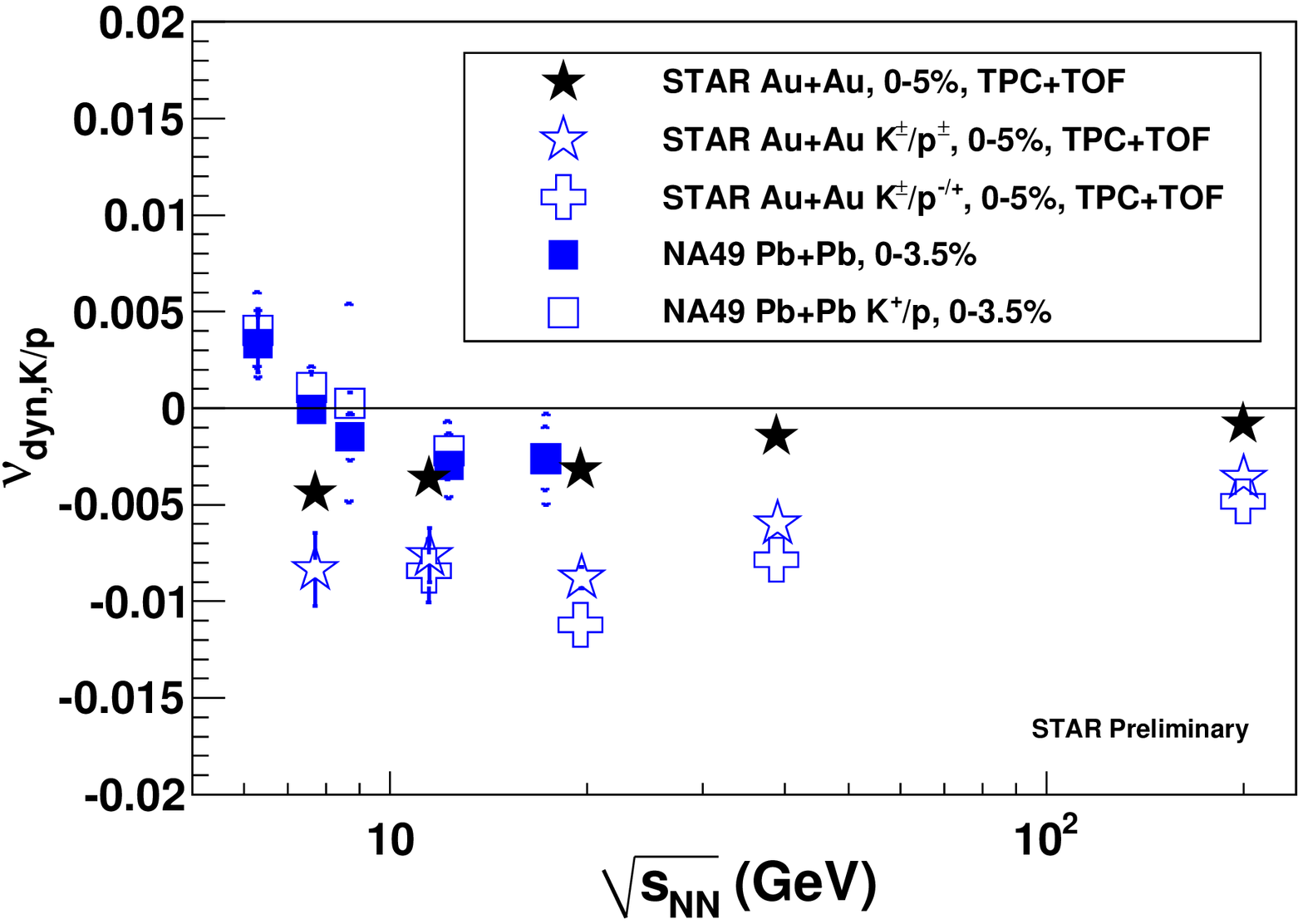} 
\label{nudyn_kp_excitation}
}
\caption{Results for the measurement of inclusive charged (closed stars) and average same-sign (open stars) and opposite-sign (open crosses), charge dependent $\nu_{\rm dyn,p/\pi}$ (left) and $\nu_{\rm dyn,K/p}$ (right).}
\end{figure}

Current results on the energy dependence of measured dynamical particle ratio fluctuations from the STAR experiment are shown in Figures \ref{nudyn_kpi_excitation}, \ref{nudyn_ppi_excitation}, and \ref{nudyn_kp_excitation}. These include previously presented results from the inclusive charged $K/\pi$, $p/\pi$, and $K/p$ ratios \cite{QM_Proceeding}, as well as initial results on the charge dependence of these dynamical ratio fluctuations as a function of collision energy in the search for the QCD critical point at RHIC. Results from previous measurements of $K/\pi$ fluctuations at $\sqrt{s_{NN}}$ = 62.4 and 200 GeV for both charge dependent and independent cases are discussed in \cite{starkpiprl}.

Figure \ref{nudyn_kpi_excitation} shows $\nu_{\rm dyn,K/\pi}$ as a function of energy from the STAR experiment, measured in central 0-5\% Au+Au collisions at the energies $\sqrt{s_{NN}}$ = 7.7, 11.5, 19.6, 39, and 200 GeV. The total dynamical $K/\pi$ fluctuations (closed stars) are constructed as $K/\pi = (K^{+}+K^{-})/(\pi^{+}+\pi^{-})$. Dynamical fluctuations can also be constructed for different charge combinations. The combinations include: $++$, $--$, $+-$, and $-+$. The average of the same sign (open stars) and opposite sign (open crosses) $K/\pi$ fluctuations are also shown in Figure \ref{nudyn_kpi_excitation}. The charge dependent fluctuations provide a more detailed picture of how particle production mechanisms affect the dynamical fluctuations. Also shown are published measurements of $\sigma_{\rm dyn,K/\pi}$ from central 0-3.5\% Pb+Pb collisions from the NA49 experiment \cite{NA49_kpi_ppi}, converted to $\nu_{\rm dyn,K/\pi}$ using $\sigma_{\rm dyn}^{2} \approx \nu_{\rm dyn}$. This was also done for $p/\pi$ and $K/p$ $\sigma_{\rm dyn}$ results from NA49. STAR has also directly calculated $\sigma_{\rm dyn,K/\pi}$ and experimentally verified the relationship between $\nu_{\rm dyn}$ and $\sigma_{\rm dyn}$. The total charged dynamical $K/\pi$ fluctuations measured by both STAR and NA49 are positive. STAR observes no large change in dynamical $K/\pi$ fluctuations between the energies from $\sqrt{s_{NN}}$ = 7.7-200 GeV. 

The charge dependent dynamical $K/\pi$ fluctuations measured by STAR are negative at all energies for both same and opposite sign charge combinations. Both charged sign combinations are close in magnitude at $\sqrt{s_{NN}}$ = 200 GeV, but the average opposite sign dynamical fluctuations become more negative with decreasing energy, faster, than the same sign fluctuations. Fluctuations in opposite-sign charge combinations tend to be more negative overall than their same-sign counterparts due to stronger cross-correlation (immediate local charge conservation) of two opposite sign particles produced by a single resonance decay. 
Additionally, the charge independent fluctuations are not a simple sum of the charge dependent components. This indicates that while the charge dependent results are always negative, the charge independent results can be positive.

%
Figure \ref{nudyn_ppi_excitation} shows $\nu_{\rm dyn,p/\pi}$ as a function of energy from the STAR experiment, measured in central 0-5\% Au+Au collisions at the energies $\sqrt{s_{NN}}$ = 7.7, 11.5, 19.6, 39, and 200 GeV. The total dynamical $p/\pi$ fluctuations (closed stars) are constructed as $p/\pi = (p^{+}+p^{-})/(\pi^{+}+\pi^{-})$. As discussed for $K/\pi$ fluctuations, dynamical fluctuations for $p/\pi$ can also be constructed for different charge combinations. The average of the same sign (open stars) and opposite sign (open crosses) $p/\pi$ fluctuations are also included in Figure \ref{nudyn_ppi_excitation}. Also included are published measurements from the NA49 experiment \cite{NA49_kpi_ppi}. 

The total charged dynamical $p/\pi$ fluctuations measured by both STAR and NA49 are negative and become more negative with decreasing collision energy. The two experiments measure coincident values for dynamical $p/\pi$ fluctuations at energies below $\sqrt{s_{NN}}$ = 19.6 GeV.

Unlike $K/\pi$ fluctuations, both charge independent and charge dependent dynamical $p/\pi$ fluctuations are always negative. This indicates a strong cross-correlation exists between produced protons and pions at all energies. The strong decay of a $\Delta$ baryon is a large contributor to this cross-correlation. Examples of this decay that produce cross-correlations would be $\Delta^{++} \rightarrow p^{+} + \pi^{+}$ or $\Delta^{0} \rightarrow p^{+} + \pi^{-}$.

The charge dependent results for dynamical $p/\pi$ fluctuations become more negative at about the same rate with decreasing collision energy, until below $\sqrt{s_{NN}}$ = 19.6 GeV. The anti-proton yield drops dramatically below this energy. The measured $\overline{p}/p$ ratio decreases by $\approx$ an order of magnitude between $\sqrt{s_{NN}}$ = 17.3 and 7.6, from 0.1 to 0.01 \cite{NA49_pbarp}. Additional study is required of fluctuations involving anti-protons at the lowest energies of $\sqrt{s_{NN}}$ = 7.7 and 11.5 GeV.

%
The measurement of $\nu_{\rm dyn,K/p}$ as a function of energy from the STAR experiment, measured in central 0-5\% Au+Au collisions at the energies $\sqrt{s_{NN}}$ = 7.7, 11.5, 19.6, 39, and 200 GeV is shown in Figure \ref{nudyn_kp_excitation}. The total dynamical $K/p$ fluctuations (closed stars) are constructed as $K/p = (K^{+}+K^{-})/(p^{+}+p^{-})$. Different charge combinations of dynamical $K/p$ fluctuations are also examined. The average of the same sign (open stars) and opposite sign (open crosses) $K/p$ fluctuations are presented in Figure \ref{nudyn_kp_excitation}. Also included are published measurements from the NA49 experiment \cite{NA49_kp}. 
Similar to $p/\pi$ fluctuations, both charge independent and charge dependent dynamical $K/p$ fluctuations are always negative. This indicates a strong cross-correlation exists between produced kaons and protons at all energies. The charge independent dynamical $K/p$ fluctuations become more negative as the collision energy is decreased. Charge dependent dynamical $K/p$ fluctuations become more negative from $\sqrt{s_{NN}}$ = 200 to 19.6 GeV, below which the fluctuations are almost constant. Results from the NA49 experiment for both charge independent and charge dependent $K/p$ fluctuations demonstrate a rapid increase below energies of $\sqrt{s_{NN}}$ = 8 GeV and actually cross zero and become positive at the lowest energies measured. The results from STAR are consistent with negative dynamical $K/p$ fluctuations at all energies. However, as for dynamical $p/\pi$ fluctuations, the $K/p$ fluctuations involving anti-protons at the two lowest energies measured by STAR are still under study.


As defined, $\nu_{\rm dyn}$ is an extensive variable, by virtue of an explicit system-size (multiplicity) dependence. Therefore, as the system-size increases, $\nu_{\rm dyn}$ should decrease (toward zero). This is seen in the centrality dependence of $\nu_{\rm dyn}$, where it approaches zero from peripheral to central collisions where the multiplicity increases \cite{starkpiprl}. This also accounts for part of the overall trends in the energy dependence of $\nu_{\rm dyn}$. To create an intensive variable, $\nu_{\rm dyn}$ scaled by the system-size (particle multiplicity) can be studied. In \cite{starkpiprl}, the scaling used are values of $dN/d\eta$ for each centrality, corrected for detector efficiency and acceptance. For this study, the uncorrected values of $dN/d\eta$ from 0-5\% central Au+Au collisions at each energy are used.

\begin{figure}[]
\advance\leftskip-2cm
\subfigure[]{
\includegraphics[width=0.64\textwidth]{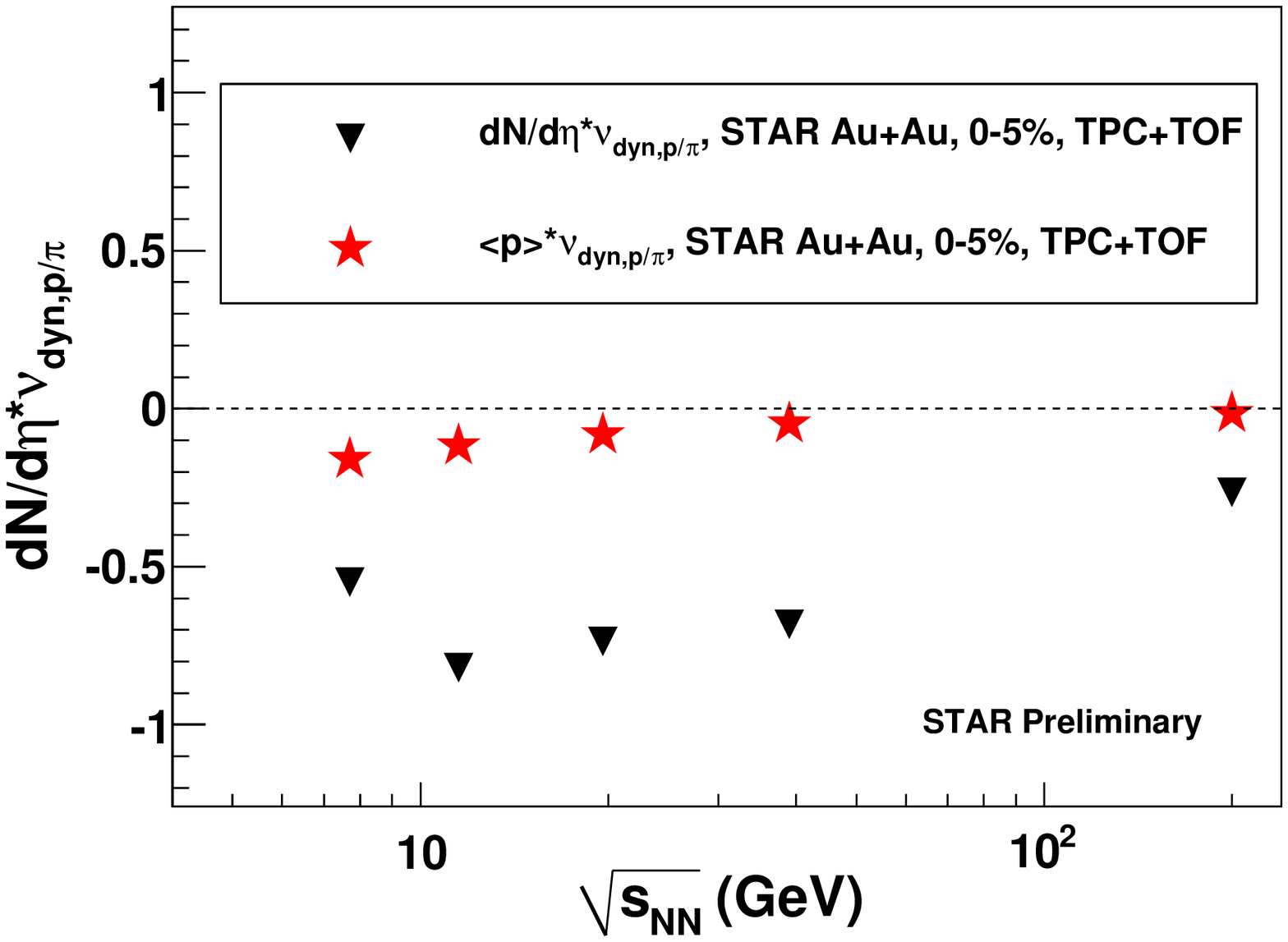} 
\label{scaled_ppi_excitation}
}
\subfigure[]{
\includegraphics[width=0.64\textwidth]{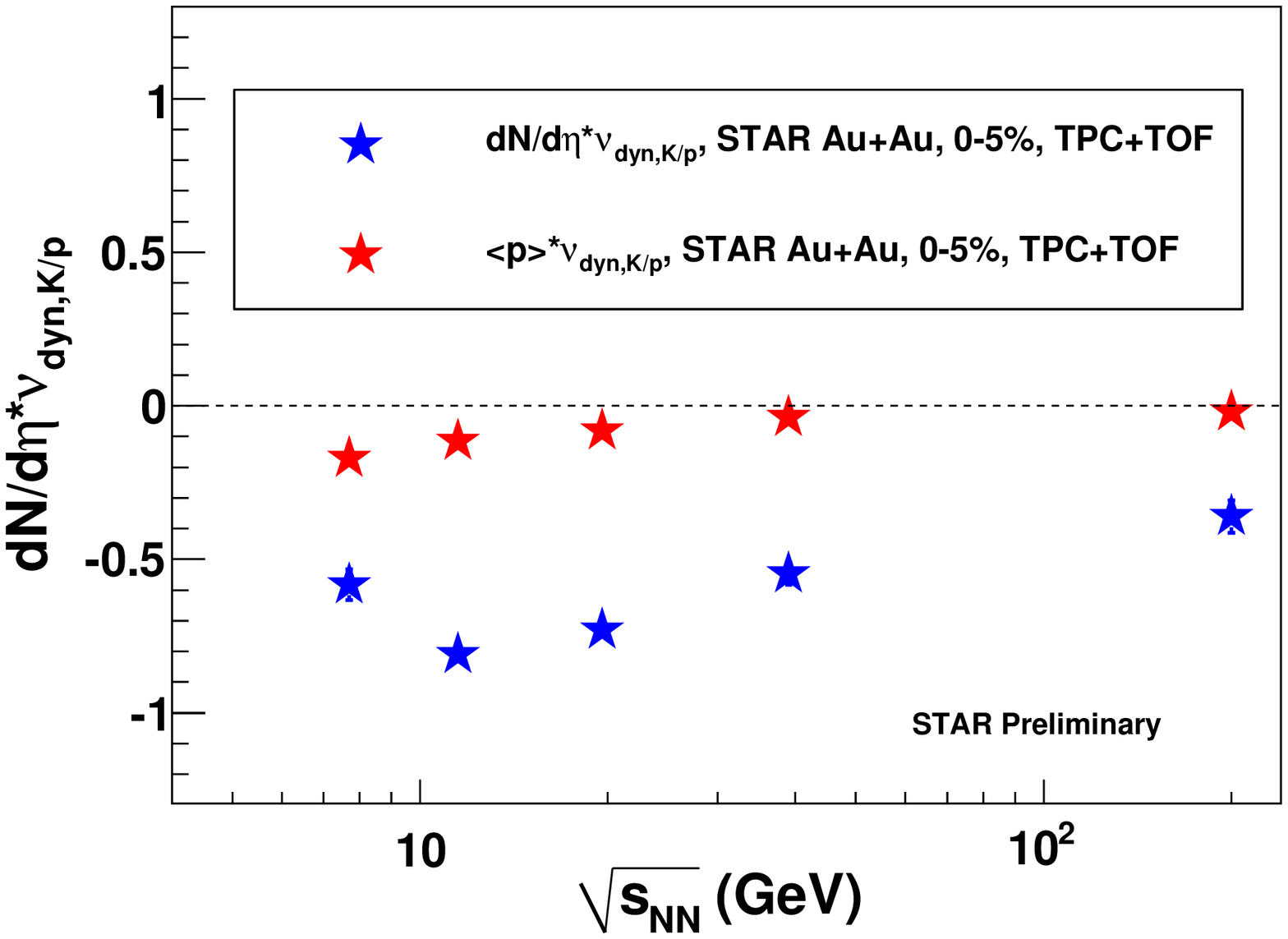} 
\label{scaled_kp_excitation}
}
\caption{Results for the measurement of $\nu_{\rm dyn,p/\pi}$ (left) and $\nu_{\rm dyn,K/p}$ (right) scaled by uncorrected charged particle multiplicity, $dN/d\eta$, from central 0-5\% Au+Au collisions at $\sqrt{s_{NN}}$ = 7.7-200 GeV.}
\end{figure}

Figures \ref{scaled_ppi_excitation}, and \ref{scaled_kp_excitation} show the values of $\nu_{\rm dyn,K/\pi}$, $\nu_{\rm dyn,p/\pi}$, and $\nu_{\rm dyn,K/p}$, respectively, scaled by uncorrected charged particle multiplicity $dN/d\eta$. Figures \ref{scaled_ppi_excitation} and \ref{scaled_kp_excitation} also include the scaling result if average uncorrected number of protons $+$ anti-protons ($<p>$) is used. Figures \ref{scaled_ppi_excitation} and \ref{scaled_kp_excitation} indicates that the uncorrected $dN/d\eta$ scaled fluctuations involving protons reach a minimum value between $\sqrt{s_{NN}}$ = 7.7 and 19.6 GeV, before increasing toward zero at higher energies. Since the net-baryon density ($\mu_{B}$) also changes with energy, further investigation of this behavior is required. The smooth scaling with only $<p>$ could reflect the change in net-baryon number with energy, as the ratio of $\mu_{B}$ at $\sqrt{s_{NN}}$ = 7.7 and 200 is similar to the ratio of $<p>*\nu_{\rm dyn,p/\pi}$ at those energies.

%
%
%
\section{Summary}

Results from dynamical particle ratio fluctuations ($K/\pi$, $p/\pi$, and $K/p$) and new results on the charge dependence of dynamical fluctuations have been presented from data acquired in Au+Au collisions at energies from $\sqrt{s_{NN}}$ = 7.7-200 GeV. Also discussed are initial results from multiplicity scaled dynamical fluctuations of all three particle ratios at energies from $\sqrt{s_{NN}}$ = 7.7-200 GeV. 
The charge dependent dynamical fluctuations provide additional insight into particle production as a function of energy. Further study is necessary at all energies, including plotting each charge combination separately. 
Uncorrected multiplicity scaled dynamical $p/\pi$ and $K/p$ fluctuations become more negative between $\sqrt{s_{NN}}$ = 7.7 and 19.6 GeV, before increasing toward zero at higher energies. 
Additional data points at $\sqrt{s_{NN}}$ = 27 and 62.4 GeV are under study and will complete the excitation function for the first phase of the RHIC energy scan to search for the QCD critical point.

\bigskip 

\end{document}